 \documentclass[aps,pra,superscriptaddress,amsmath,amssymb,preprintnumbers,twocolumn,floatfix,showpacs,showkeys,10pt]{revtex4-1}
 \usepackage{amssymb} \usepackage{epsfig}
 \begin{document}
  \title{Entropic uncertainty relation under quantum channels with memory  }

\author{You-neng Guo}
\email{guoxuyan2007@163.com}
\affiliation{College of Electronic and Communication Engineering, Changsha University, Changsha, Hunan
410022, People's Republic of China}
\affiliation{Key Laboratory of Low-Dimensional Quantum Structures and
Quantum Control of Ministry of Education, and Department of Physics,
Hunan Normal University, Changsha 410081, People's Republic of
China}
\author{Mao-fa Fang}
\email{mffang@hunnu.cn}
\affiliation{Key Laboratory of Low-Dimensional Quantum Structures and
Quantum Control of Ministry of Education, and Department of Physics,
Hunan Normal University, Changsha 410081, People's Republic of
China}
\author{Ke Zeng}
\email{zk92@126.com}
\affiliation{College of Electronic and Communication Engineering, Changsha University, Changsha, Hunan
410022, People's Republic of China}

\begin{abstract}
Recently, Xu et al. [Phys. Rev. A 86, 012113(2012)] explored the behavior of the entropic uncertainty relation under the influence of local unital and nonunital noisy channels for a class
of Bell-diagonal states. We here reform their results and investigate the entropic uncertainty relation under the influence of unital and nonunital noisy channels with memory. Different types of noisy channels with memory, such as amplitude damping channel(nonunitary), phase-damping and depolarizing channels(unitary) have been taken into account. Some analytical or numerical results are presented. The effect of channels with memory on dynamics of the entropic uncertainties (or their lower bounds) has been discussed in detail. Compare with previous results, our results show that, the entropic uncertainties (or their lower bounds) subjecting to amplitude damping channel with memory will be reduced at first and then be lifted with the memory coefficient of channel $\mu$ increasing, however they will be only reduced under phase-damping and depolarizing channels with memory. Especially, in the limit of $\mu\rightarrow1$, the entropic uncertainties (or their lower bounds) could be well protected and immune to decoherence of channle. Moreover, the mechanism behind these phenomena are also explored by using the purity of state.

 \end{abstract}

  \pacs{73.63.Nm, 03.67.Hx, 03.65.Ud, 85.35.Be}
 \maketitle
\section{Introduction}
The uncertainty principle is one of the most remarkable features of quantum mechanics initially proposed by Heisenberg~\cite{Heisenberg} for two incompatible observables of position and momentum. It states that, one can't simultaneously predict the measurement outcomes of these two incompatible observables with certainty. Later this uncertainty principle is further formulated by Robertson~\cite{Robertson} to generalize for arbitrary pairs of observables $R$ and $Q$ based on the standard deviations: $\Delta R\cdot\Delta Q\geq\frac{1}{2}|\langle\psi[R,Q]\psi\rangle|$, and $\Delta R(\Delta Q)$ are the standard deviations and $[R,Q]=RQ-QR$ is the commutator. Usually, the left hand side of this inequality is named uncertainty and the right hand of this inequality is called uncertainty bound. It is worthy noting that, this uncertainty bound is dependent of quantum state $|\psi\rangle$ to be measured, which maybe leads to a trivial bound if the commutator has zero expectation value. To break through this drawback and to precisely capture its physical meanings, uncertainty relation has been developed in an information theoretical framework replacing the standard deviation with entropy by Deutsch~\cite{Deutsch}. Afterwards, an improved version was given by Kraus and then proved by Maassen and Uiffink~\cite{Maassen}. It states that, given arbitrary pairs of observables $R$ and $Q$ with eigenbases $|\varphi_{i}\rangle$ and $|\psi_{j}\rangle$, respectively, for any state $\rho$ to be measured
\begin{equation}\label{Eq1}
H(R)+H(Q)\geq\log_{2}\frac{1}{c}
\end{equation}
where $c=\max_{i,j}|\langle\varphi|\psi\rangle|^2$ quantifies the complementarity of $R$ and $Q$, and $H(R)$ or $H(Q)$ denotes the Shannon entropy of the probability
distribution of the outcomes when $R$ or $Q$ is performed on a system $\rho$, respectively. Note that this uncertainty bound quantified by entropy provides a fixed lower bound comparing with Heisenberg's uncertainty relation, since $c$ does not depend on specific states to be measured.

However, under certain circumstance, the above uncertainty relation may be violated if an observer prepares a maximal correlated bipartite state $\rho_{AB}$ and sends just one of the particles to another one and keeps the other particle as a quantum memory by himself, then he is able to precisely predict the outcomes. This is the quantum-memory-assisted entropic uncertainty principle, which was initially proposed by Renes and Boileau~\cite{Renes,Boileau} and subsequently was proven by Berta et al.~\cite{Berta}, as well as confirmed by several experimentally~\cite{Prevedel,Li}. This new relation can read as
\begin{equation}\label{Eq2}
S(R|B)+S(Q|B)\geq\log_{2}\frac{1}{c}+S(A|B)
\end{equation}
where $S(X|B)$ is the conditional von Neumann entropy of the post measurement states
\begin{equation}\label{Eq3}
\rho_{XB}=\sum_{X}(|\phi_{X}\rangle\langle\phi_{X}|\otimes I)\rho_{AB}(|\phi_{X}\rangle\langle\phi_{X}|\otimes I)
\end{equation}
where $|\phi_{X}\rangle$ are the eigenvectors of $X\in(R,Q)$. However, comparing with the previous uncertainty relation, this new uncertainty bound can be reduced if the state $\rho_{AB}$ is initially entangled. On the other hand, if the state $\rho_{AB}$ is initially unentangled, the above new uncertainty relation reduces to Eq.(1). Recently, the improvement of this new entropic uncertainty principle as well as its dynamics have attracted increasing attention~\cite{Tomamichel,Coles,Hu,Feng,Pati,Adabi,Adabi1,Wang,Huang1} and have its potential applications such as for witnessing entanglement~\cite{Xu,Partovi}, teleportation~\cite{Hu1} and cryptography~\cite{Ng}.

In this paper, we explore the behavior of this new entropic uncertainty relation under the influence of noisy channels in the presence of memory. As we known any quantum systems inevitably suffer from decoherence or dissipation due to the interaction between the systems and the environment. Therefore, it is important to investigate how noisy environment influences on the entropic uncertainty. Recently, Xu et al.~\cite{Xu1} firstly studied the influence of local unital and nonunital noisy channels on the quantum-memory-assisted entropic uncertainty relation only for a class of Bell-diagonal states, they concluded that the amount of uncertainty increases in the present of the unital noises, but may reduce with the amplitude-damping nonunital noises. Later, their results are confirmed by Huang et al.~\cite{Huang1} when they explored the dynamics of the uncertainty for three different kinds of Pauli-observable measurement under local unitary and nonunitary channels. We here go on to investigate the entropic uncertainty relation under the noisy channels with memory. We mainly focus on how the unital and nonunital noisy channels with memory influence on the entropic uncertainty relation. Different types of noisy channels with memory, such as amplitude damping channel(nonunitary), phase-damping and depolarizing channels(unitary) have been taken into account. Comparing with noisy channels without memory, the effect of channels with memory on dynamics of the entropic uncertainties (or their lower bounds) has been discussed in detail. Some analytical or numerical results are presented. The results show that, the dynamics of the uncertainty relation is strongly related to channels with memory. The entropic uncertainties (or their lower bounds) subjecting to amplitude damping channel with memory will be reduced at first and then be lifted with the memory coefficient of channel $\mu$ increasing. However,  phase-damping and depolarizing channels with memory can reduce the entropic uncertainties (or their lower bounds). The stronger the memory coefficient of channel $\mu$ is, the lower the amount of entropic uncertainties are. Especially, in the limit of $\mu\rightarrow1$, the entropic uncertainties (or their lower bounds) could be well protected and immune to decoherence of channle.

The layout is as follows: In Sec. \textrm{II}, we illustrate initial states and noise channels. In  Sec. \textrm{III}, we devote to examining dynamics behavior of the entropic uncertainties (or their lower bounds) in different types of noisy channels with memory, and adopt the purity of the state to explain the changing of
the entropic uncertainties relation. Finally, we give the conclusion  in Sec. \textrm{IV}.

\section{Initial states and noise model}  
To begin with, we review a brief description of quantum channels with memory. Usually, there are two different types of quantum channels including memory and memoryless channels. Assume any two particles are sent at a time interval $t$, and $\tau$ denotes the characteristic memory timescale for the noisy channels. When the noisy environmental correlation time is smaller than the time between consecutive
uses, namely $\tau < t$ at each channel use the environment back action can be negligible. Define $\varepsilon$ stands for a quantum noisy operation process is independent noise acts on each uses. Therefore, the dynamics map of input-state
to output-state density matrices can be described by $\varepsilon_{N} = \varepsilon_{}^{\otimes N}(\rho)$, here $N$ represents times uses of channel. However, when the correlation times of noisy channels are longer than the time between consecutive uses, namely $\tau > t$ corresponds to the noisy channels with memory so that the channel acts dependently on each channel input and the dynamics map $\varepsilon_{N}\neq\varepsilon_{}^{\otimes N}$.
In what follows we consider two times consecutive uses channel. Based on the Kraus operator approach, for any initial state $\rho$, the evolution of initial state under noise is given by~\cite{Macchiavello,Yeo}
\begin{equation}\label{Eq3}
\varepsilon(\rho)=(1-\mu)\sum_{i,j}E_{i,j}^{u}\rho E_{i,j}^{u\dagger}+\mu\sum_{k}E_{k,k}^{c}\rho E_{k,k}^{c\dagger}
\end{equation}
The above expression indicates that the same operation with probability $\mu$ specified
by $E_{k,k}^{c}$ is correlated noise while with probability $1-\mu$ corresponding to $E_{i,j}^{u}$ is uncorrelated noise. In the following, we make use of Eq.~\eqref{Eq3} to study the effect of memory parameter of quantum channels on entropic uncertainty relation. In order to investigate the behaviour of the uncertainty relation under the influence
of noisy channels with memory, in what follows we will consider a observer
possesses two particles (A) and (B) suffer a noisy channel with memory and sends a particle (A) to another observer, who carries out one of the measurements $R$ and $Q$ on his particle
(A), and then he informs his measurement. Without loss of generality, we assume two particles states are initially in
a general class of two-qubit state which can be written in the Hilbert-Schmidt representation as
\begin{equation}\label{A0}
\rho=\frac{1}{4}(I\otimes I+\sum_{i}^{i=3}a_{i}\sigma_{i}\otimes I+I \otimes\sum_{i}^{i=3}b_{i}\sigma_{i}+\sum_{i,j}^{i,j=3}c_{ij}\sigma_{i}\otimes \sigma_{j}),
\end{equation}
here $I$ stands for $2\otimes2$ the identity operator, $\sigma_{i}$ are the standard Pauli matrices. For an arbitrary two-qubit state, one can use these parameters $(a_{i}, b_{i}, c_{ij})$ to represent them. However, the general state of two-qubit system described by Eq.~\eqref{A0}, under proper unitary rotations~\cite{Horodecki,Akhtarshenas},  can be parameterized by nine real parameters $\vec{X}=(x_{1},x_{2},x_{3})^{t}$, $\vec{Y}=(y_{1},y_{2},y_{3})^{t}$ and the correlation matrix $T=(t_{11},t_{22},t_{33})$.

\begin{widetext}
\makeatletter\def\@captype{table}\makeatother
\caption{Kraus operators for amplitude damping, phase damping, and depolarizing channels with memory, where $D$ represents the decoherence parameter.}$
\begin{tabular}{|c|c|c|c|}
\hline
$\text{Channel description}$&$\text{ Uncorrelated Kraus operators}$&$\text{Correlated Kraus operators}$\\
\hline
$\text{Amplitude damping channel}$ & $
\begin{tabular}{l} $E_{ij}^{u}=A_{i}\otimes A_{j},\quad (i,j=0,1)$ \\ $A_{0}=\left[
\begin{array}{cc}
\sqrt{1-D} & 0 \\
0 & 1%
\end{array}%
\right] ,$ $A_{1}=\left[
\begin{array}{cc}
0 & 0 \\
\sqrt{D} & 0%
\end{array}%
\right] $
\end{tabular}
$ & $
E_{00}^{c}=\left[
\begin{array}{cccc}
\sqrt{1-D} & 0 & 0 & 0 \\
0 & 1 & 0 & 0 \\
0 & 0 & 1 & 0 \\
0 & 0 & 0 & 1
\end{array}%
\right] ,$ $E_{11}^{c}=\left[
\begin{array}{cccc}
0 & 0 & 0 & 0 \\
0 & 0 & 0 & 0 \\
0 & 0 & 0 & 0 \\
\sqrt{D} & 0 & 0 & 0
\end{array}%
\right]
$ \\
\hline
$\text{Phase damping channel}$ &  $
\begin{tabular}{l} $E_{ij}^{u}=\sqrt{P_{i}P_{j}}\sigma_{i}\otimes \sigma_{j},\quad (i,j=0,3)$ \\ $\sigma_{0}=\left[
\begin{array}{cc}
1 & 0 \\
0 & 1%
\end{array}%
\right] ,$ $\sigma_{3}=\left[
\begin{array}{cc}
1 & 0 \\
0 & -1%
\end{array}%
\right] $
\end{tabular}
$ &  $
\begin{tabular}{l} $E_{kk}^{c}=\sqrt{P_{k}}\sigma_{k}\otimes \sigma_{k},\quad (k=0,3)$ \\ $\sigma_{0}=\left[
\begin{array}{cc}
1 & 0 \\
0 & 1%
\end{array}%
\right] ,$ $\sigma_{3}=\left[
\begin{array}{cc}
1 & 0 \\
0 & -1%
\end{array}%
\right] $
\end{tabular}
$ \\
\hline
$\text{Depolarizing channel}$ & $
\begin{tabular}{l} $E_{ij}^{u}=\sqrt{P_{i}P_{j}}\sigma_{i}\otimes \sigma_{j},\quad (i,j=0,1,2,3)$ \\ $\sigma_{0}=\left[
\begin{array}{cc}
1 & 0 \\
0 & 1%
\end{array}%
\right] ,$ $\sigma_{1}=\left[
\begin{array}{cc}
0 & 1 \\
1 & 0%
\end{array}%
\right] $ \\ $\sigma_{2}=\left[
\begin{array}{cc}
0 & -i \\
i & 0%
\end{array}%
\right] ,$ $\sigma_{3}=\left[
\begin{array}{cc}
1 & 0 \\
0 & -1%
\end{array}%
\right] $
\end{tabular}
$ &  $
\begin{tabular}{l} $E_{kk}^{c}=\sqrt{P_{k}}\sigma_{k}\otimes \sigma_{k},\quad (k=0,1,2,3)$ \\ $\sigma_{0}=\left[
\begin{array}{cc}
1 & 0 \\
0 & 1%
\end{array}%
\right] ,$ $\sigma_{1}=\left[
\begin{array}{cc}
0 & 1 \\
1 & 0%
\end{array}%
\right] $\\ $\sigma_{2}=\left[
\begin{array}{cc}
0 & -i \\
i & 0%
\end{array}%
\right] ,$ $\sigma_{3}=\left[
\begin{array}{cc}
1& 0 \\
0 & -1%
\end{array}%
\right] $
\end{tabular}
$ \\
\hline
\end{tabular}%
$%

\begin{table}[tbh]
\newcommand{\tabincell}[2]{\begin{tabular}{@{}#1@{}}#2\end{tabular}}
\caption{The evolved states of the initial density matrix parameters under amplitude damping, phase damping and depolarizing channels with memory}$%
\begin{tabular}{|c|c|c|c|c|}
\hline
$\text{$\varepsilon[\rho(0)]$}$ & $\text{Bloch vector $\vec{X}=(x1,x2,x3)^{T}$}$ & $\text{Bloch vector $\vec{Y}=(y1,y2,y3)^{T}$}$ & $\text{Correlation matrix $T=t_{ij}$}$\\
\hline
$\text{Am}$ & \begin{tabular}{l} $x1=\frac{1}{2}(2\sqrt{1-D}+\mu-\sqrt{1-D}\mu)a_{1}$ \\
$x2=\frac{1}{2}(2\sqrt{1-D}+\mu-\sqrt{1-D}\mu)a_{2}$ \\
\tabincell{c}{$x3=\frac{1}{2}(1+a_{3}-b_{3}-c_{3})D\mu$\\
$-D+(1-D)a_{3}$}
\end{tabular}& $
\begin{tabular}{l}
$y1=\frac{1}{2}(2\sqrt{1-D}+\mu-\sqrt{1-D}\mu)b_{1}$ \\
$y2=\frac{1}{2}(2\sqrt{1-D}+\mu-\sqrt{1-D}\mu)b_{2}$ \\
\tabincell{c}{$y3=\frac{1}{2}(1-a_{3}+b_{3}-c_{3})D\mu$\\
$-D+(1-D)b_{3}$}
\end{tabular}
$ & $ $
\begin{tabular}{l}
$t_{11}=\frac{c_{1}}{2}(1-\sqrt{1-D})\mu(c_{2}-c_{1})+(1-D-D\mu)$\\
$t_{22}=\frac{c_{2}}{2}(1-\sqrt{1-D})\mu(c_{1}-c_{2})+(1-D-D\mu)$\\
\tabincell{c}{$t_{33}=D(1-\mu)(1-D)(c_{3}-a_{3}-b_{3})$\\
$+D^2(1-\mu)+c_{3}$}\\
$t_{13}=\frac{1}{2}\mu(\sqrt{1-D}-1)a_{1}-D\sqrt{1-D}(1-\mu)a_{1}$\\
$t_{23}=\frac{1}{2}\mu(\sqrt{1-D}-1)a_{2}-D\sqrt{1-D}(1-\mu)a_{2}$\\
$t_{31}=\frac{1}{2}\mu(\sqrt{1-D}-1)b_{1}-D\sqrt{1-D}(1-\mu)b_{1}$\\
$t_{32}=\frac{1}{2}\mu(\sqrt{1-D}-1)b_{2}-D\sqrt{1-D}(1-\mu)b_{2}$\\
$t_{12}=t_{21}=0$
\end{tabular}
$$\\
\hline
$\text{Ph}$ & \begin{tabular}{l} $x1=(1-D)a_{1}$ \\
$x2=(1-D)a_{2}$ \\
$x3=a_{3}$
\end{tabular}& $
\begin{tabular}{l}
$y1=(1-D)b_{1}$ \\
$y2=(1-D)b_{2}$ \\
$y3=b_{3}$
\end{tabular}
$ & $ $
\begin{tabular}{l}
$t_{11}=c_{1}[1-(1-D)D(1-\mu)]$\\
$t_{22}=c_{2}[1-(1-D)D(1-\mu)]$\\
$t_{33}=c_{3}$\\
$t_{12}=t_{21}=t_{23}=t_{32}=t_{13}=t_{31}=0$
\end{tabular}
$$\\
\hline
$\text{De}$ & \begin{tabular}{l} $x1=(1-\frac{4D}{3})a_{1}$ \\
$x2=(1-\frac{4D}{3})a_{2}$ \\
$x3=(1-\frac{4D}{3})a_{3}$
\end{tabular}& $
\begin{tabular}{l}
$y1=(1-\frac{4D}{3})b_{1}$ \\
$y2=(1-\frac{4D}{3})b_{2}$ \\
$y3=(1-\frac{4D}{3})b_{3}$
\end{tabular}
$ & $ $
\begin{tabular}{l}
$t_{11}=\frac{1}{9}c_{1}[9-8D(3-2D)(1-\mu)]$\\
$t_{22}=\frac{1}{9}c_{2}[9-8D(3-2D)(1-\mu)]$\\
$t_{33}=\frac{1}{9}c_{3}[9-8D(3-2D)(1-\mu)]$\\
$t_{12}=t_{21}=t_{23}=t_{32}=t_{13}=t_{31}=0$
\end{tabular}
$$\\
\hline
\end{tabular}%
$%
\end{table}
\end{widetext}

\section{Dynamics of the entropic uncertainties or their lower bounds under quantum channels with memory }  
In this section, we will probe the dynamics of the entropic uncertainty relation in different types of noisy channels with memory, such as amplitude damping channel, phase-damping and depolarizing channels. How do the channels with memory affect on dynamics of the entropic uncertainties (or their lower bounds) has been discussed in detail.

\subsection{Amplitude damping channel with memory}
Amplitude damping channel which is used to characterize spontaneous emission represents the dissipative interaction between system and
the environment. The Kraus operators for the amplitude damping channel in the presence of memory is given in Table I. According to Eq. (3), the evolved states of the initial density matrix of such a
system when it is influenced by the amplitude damping channel with memory is also given in Table II. If one chooses two of the Pauli observables $R=\sigma_{x}$ or $Q=\sigma_{z}$, in the computational
basis$\{|00\rangle,|01\rangle,|10\rangle,|11\rangle\}$, the post measurement states are
\begin{widetext}
\begin{eqnarray}\label{Eq7}
\rho_{\sigma_{x}B}^{Am}&=&\frac{1}{8}[2(1-D)+2 b_{3}(1-D)+D\mu(1-a_{3}+b_{3}-c_{3})](|00\rangle\langle00|+|10\rangle\langle10|)\nonumber\\
&+&\frac{1}{8}[2(1+D)-2 b_{3}(1-D)-D\mu(1-a_{3}+b_{3}-c_{3})](|01\rangle\langle01|+|11\rangle\langle11|)\nonumber\\
&+&\{\frac{1}{8}[(1-\sqrt{1-D})\mu(c_{2}-c_{1})+2c_{1}(1-D+D\mu)](|00\rangle\langle11|+|01\rangle\langle10|)\nonumber\\
&+&\frac{1}{8}[2\sqrt{1-D}+(1-\sqrt{1-D})\mu](b_{1}-i b_{2})(|00\rangle\langle01|+|10\rangle\langle11|)\nonumber\\
&+&\frac{1}{4}[\sqrt{1-D}(1+D)(1-\mu)+\mu]a_{1}|01\rangle\langle11|\nonumber\\
&+&\frac{1}{4}\sqrt{1-D}[1-D(1-\mu)]a_{1}|00\rangle\langle10|+H.C.\}\\
\rho_{\sigma_{z}B}^{Am}&=&\frac{1}{4}\{1-a_{3}-b_{3}+c_{3}[(1-D)^2+(3-D)D\mu]+(D-D\mu+\mu)(1+a_{3}+b_{3})+2(1-\mu)\}|11\rangle\langle11|\nonumber\\
&+&\frac{1}{4}\{1-a_{3}+b_{3}-c_{3}[(1-D)^2+(2-D)D\mu]+D(1-\mu)[2a_{3}-D(1+a_{3}+b_{3})]\}|10\rangle\langle10|\nonumber\\
&+&\frac{1}{4}\{1+a_{3}-b_{3}-c_{3}[(1-D)^2+(2-D)D\mu]+D(1-\mu)[2b_{3}-D(1+a_{3}+b_{3})]\}|01\rangle\langle01|\nonumber\\
&+&\frac{1}{4}(1-D)(1-D+D\mu)(1+a_{3}+b_{3}+c_{3})|00\rangle\langle00|\nonumber\\
&+&\{\frac{1}{4}\sqrt{1-D}(1-D+D\mu)(b_{1}-i b_{2})|00\rangle\langle01|+H.C.\}\\
\rho_{B}^{Am}&=&\frac{1}{4}[2(1+D)-2 b_{3}(1-D)-D\mu(1-a_{3}+b_{3}-c_{3})]|1\rangle\langle1|\nonumber\\
&+&\frac{1}{4}[2(1-D)+2 b_{3}(1-D)+D\mu(1-a_{3}+b_{3}-c_{3})]|0\rangle\langle0|\nonumber\\
&+&\{\frac{1}{4}[2\sqrt{1-D}+(1-\sqrt{1-D})\mu](b_{1}-i b_{2})|0\rangle\langle1|+H.C.\}
\end{eqnarray}
\end{widetext}

\begin{figure}[htpb]
   \includegraphics[width=7.5cm]{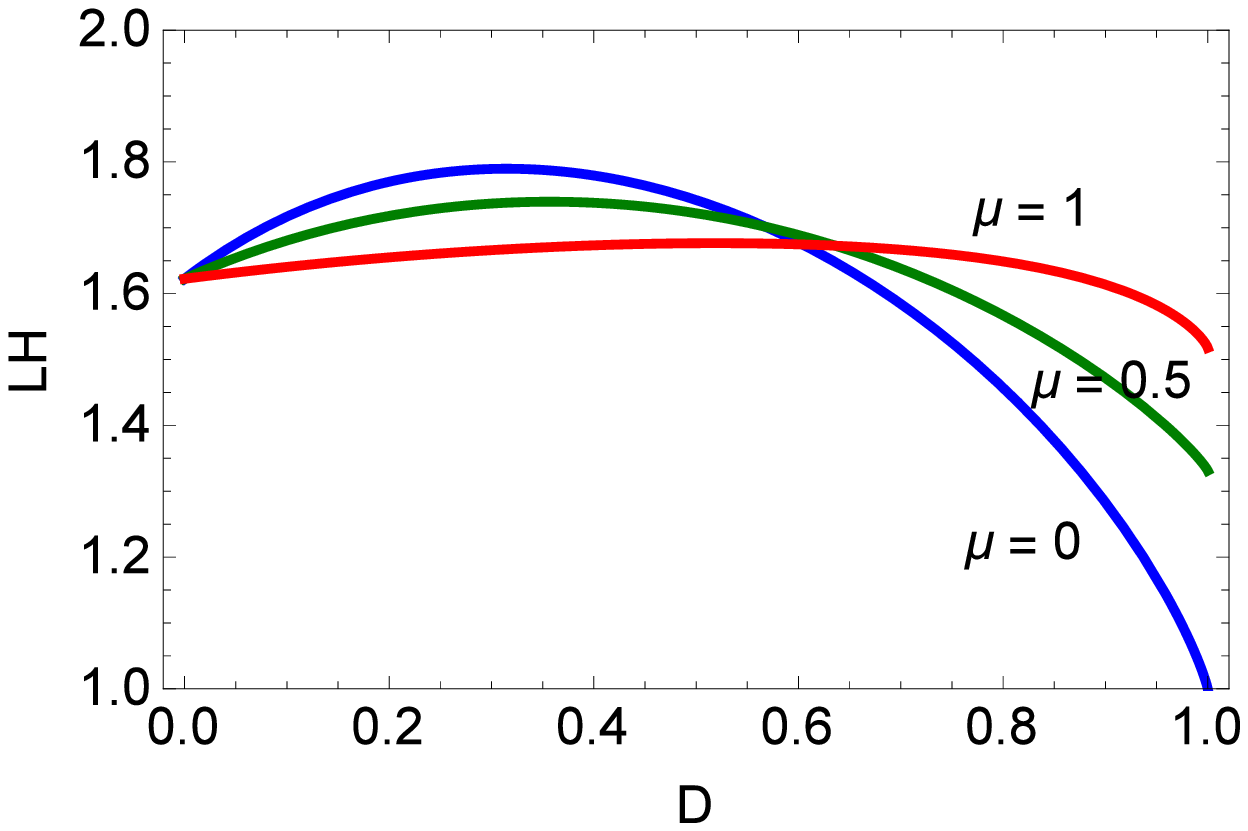}
   \includegraphics[width=7.5cm]{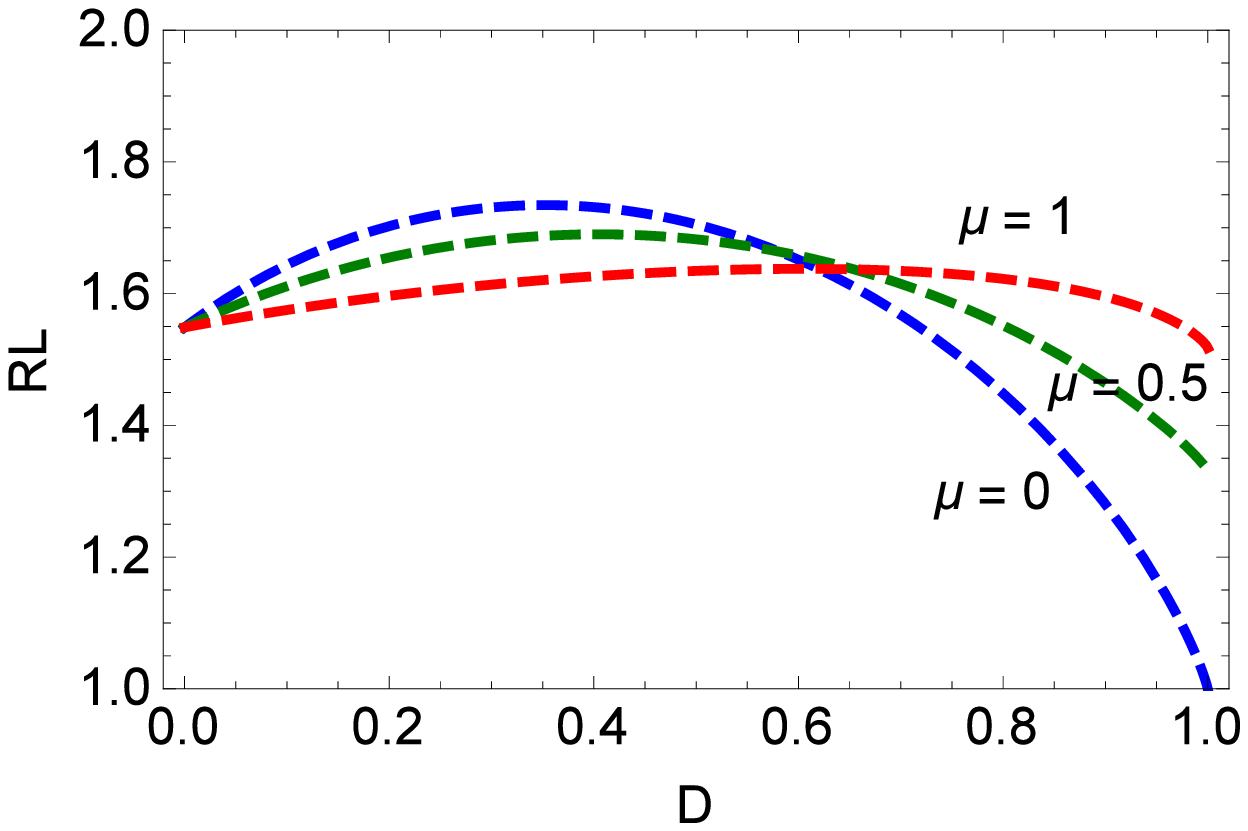}
    \includegraphics[width=7.5cm]{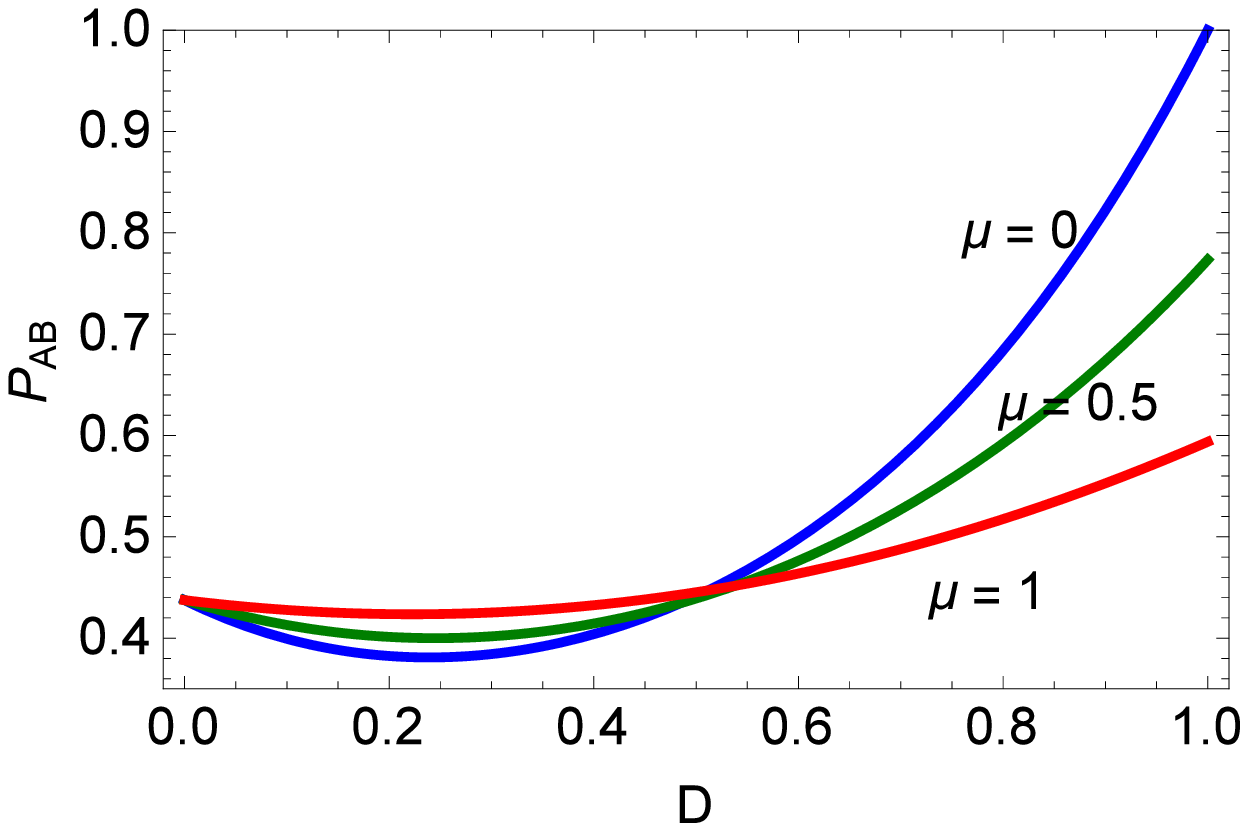}
 \caption{\label{Fig2}(Color online)The entropic uncertainties, their lower bounds and the purity of state $\rho_{AB}$ under amplitude damping channel with memory for $D$ as decoherence rate. LH represents the left-hand side of Eq. (2) and RH be the right-hand side.}
\end{figure}

The conditional von Neumann entropy of the measurement outcomes in equation (2)
would be changed into
\begin{eqnarray}\label{Eq6}
S(\sigma_{x}|B)=S(\rho_{\sigma_{x}B}^{Am})-S(\rho_{B}^{Am})\nonumber\\
S(\sigma_{z}|B)=S(\rho_{\sigma_{z}B}^{Am})-S(\rho_{B}^{Am})
\end{eqnarray}
and the right-hand side of Eq. (2) can be described as
\begin{eqnarray}\label{Eq6}
U_{b}=1+S(\rho_{AB}^{Am})-S(\rho_{B})
\end{eqnarray}

It is well known that the von Neumann entropy can be expressed as $S(\rho)=-\sum_{i}\lambda_{i}\log_{2}\lambda_{i}$, and $\lambda_{i}$ are the eigenvalues of the state of $\rho$.
In order to show directly the evolution of entropy uncertain relation, we plot the diagrams of the uncertainties and their lower bounds of the observables $\sigma_{x}$ and $\sigma_{z}$ for initial state parameters $\vec{a}=\vec{b}=(0,0,0)$ and $C=\{1/2,-1/2,1/2\}$ as a function of $D$ in Fig.1, respectively. In order to make results comparable, we have plotted
the results in amplitude damping channel both with and without memory. It is clearly seen the uncertainties will increase at first and then decrease when only subject to amplitude damping channel in the absence of memory $\mu=0$. This result is agreement with the statement made by~\cite{Xu1}.  In contrast, the behaviors of the uncertainties are more interesting in the presence of memory, with the memory coefficient of channel $\mu$ increasing, e.g.($\mu=0, 0.5, 1$), the uncertainties can be reduced with respect to weak decoherence regime, while they will be lifted in strong decoherence regime. On the other hand, their lower bounds have the same changing tendency with the uncertainties, and Eq.(2) always holds.

To get a better understanding the mechanism behind of the uncertainty, here we resort to the purity of the state which is proved to be
dramatically anti-correlated with the uncertainty relation. The purity of an evolutive state $\rho_{AB}$ is denoted as $P_{AB}=Tr(\rho_{AB}^{2})$.
To illustrate this conjecture in Fig.1, we plot $P_{AB}$ as a function of the decoherence rate $D$ in the presence of memory. From the figure, it is obvious seen that the purity of state $\rho_{AB}$ will be risen at first and then be reduced, this phenomenon is exactly anti-correlated with the uncertainty relation. Namely, the uncertainties (or their low bounds) reduce while the purity of state $\rho_{AB}$ rises, vice versa.

\subsection{Phase-damping channel with memory}
Phase damping channel which is a unital channel describes a quantum noise with loss of quantum phase information but not loss of energy. The Kraus operators for quantum dephasing channel with memory is given in Table I, where $P_{0}=1-D$, $P_{3}=D$, with $D=1-\exp(-\gamma t)$. If the particle (A) is measured by one of the Pauli operators $R=\sigma_{x}$ and $Q=\sigma_{z}$, the post measurement states are obtained

\begin{widetext}
\begin{eqnarray}\label{Eq7}
\rho_{\sigma_{x}B}^{Ph}&=&\frac{1}{4}(1+b_{3})(|00\rangle\langle00|+|10\rangle\langle10|)+\frac{1}{4}(1-b_{3})(|01\rangle\langle01|+|11\rangle\langle11|)\nonumber\\
&+&\{\frac{1}{4}c_{1}[1-D(2-D)(1-\mu)](|00\rangle\langle11|+|01\rangle\langle10|)\nonumber\\
&+&\frac{1}{4}(1-D)(b_{1}-i b_{2})(|00\rangle\langle01|+|10\rangle\langle11|)\nonumber\\
&+&\frac{1}{4}(1-D)a_{1}(|01\rangle\langle11|+|00\rangle\langle10|)+H.C.\}\\
\rho_{\sigma_{z}B}^{Ph}&=&\frac{1}{4}(1-a_{3}-b_{3}+c_{3})|11\rangle\langle11|+
\frac{1}{4}(1-a_{3}+b_{3}-c_{3})|10\rangle\langle10|\nonumber\\
&+&\frac{1}{4}(1+a_{3}-b_{3}-c_{3})|01\rangle\langle01|+
\frac{1}{4}(1+a_{3}+b_{3}+c_{3})|00\rangle\langle00|\nonumber\\
&+&\{\frac{1}{4}(1-D)(b_{1}-i b_{2})(|00\rangle\langle01|+|10\rangle\langle11|)+H.C.\}\\
\rho_{B}^{Ph}&=&\frac{1}{2}(1-b_{3})|1\rangle\langle1|+\frac{1}{2}(1+b_{3})|0\rangle\langle0|+
\{\frac{1}{2}(1-D)(b_{1}-i b_{2})|0\rangle\langle1|+H.C.\}
\end{eqnarray}
\end{widetext}

\begin{figure}[htpb]
   \includegraphics[width=7.5cm]{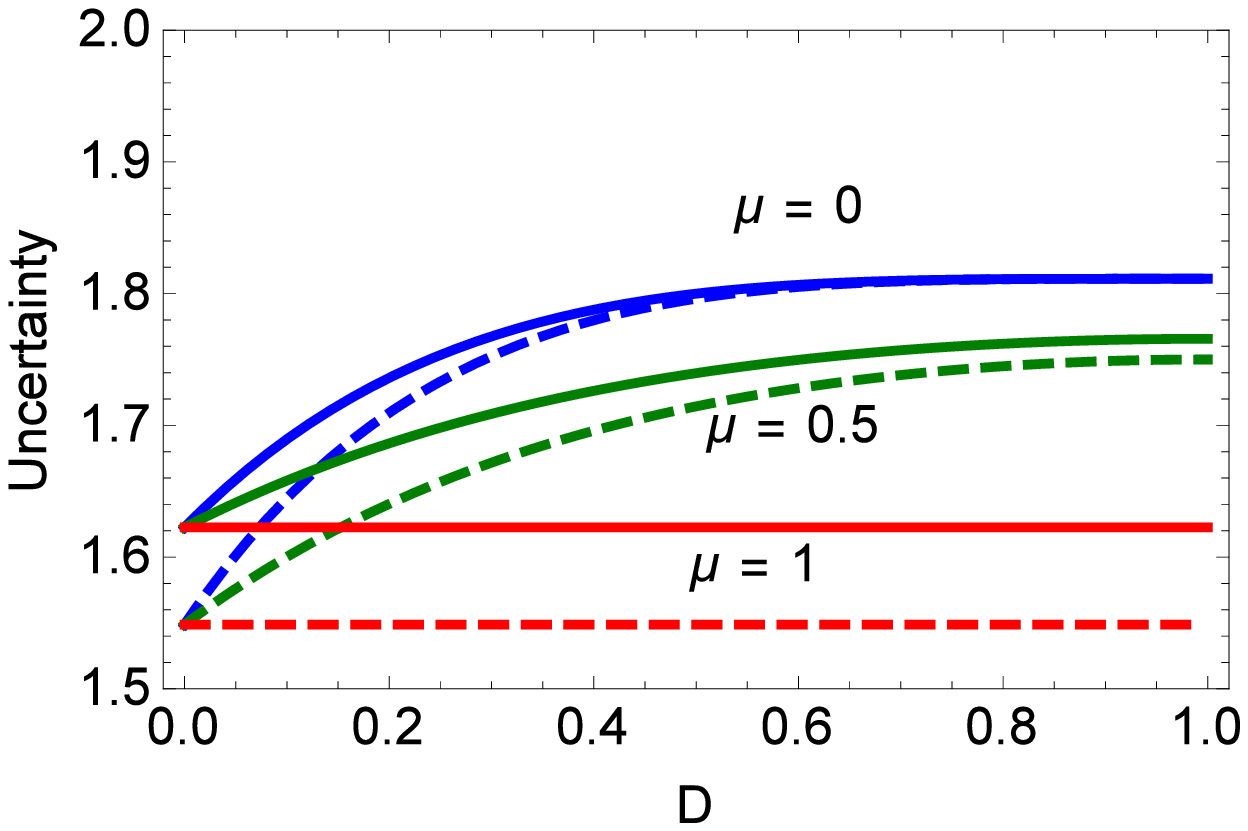}
    \includegraphics[width=7.5cm]{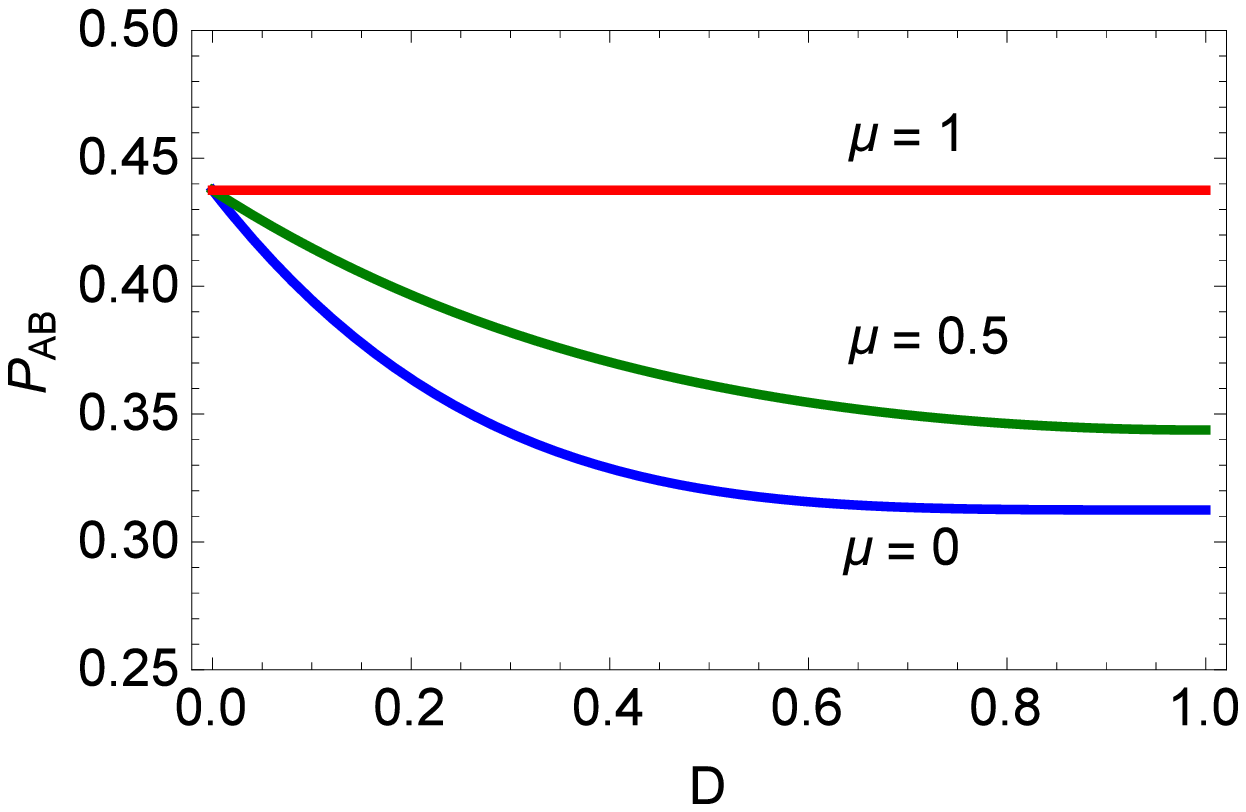}
 \caption{\label{Fig3}(Color online)The entropic uncertainties, their lower bounds and the purity of state $\rho_{AB}$ under phase damping channel with memory for $D$ as decoherence rate. The solid lines represent the entropic uncertainties and the dashed lines represent the lower bounds of uncertainties.}
\end{figure}
By replacing the $S(\rho_{\sigma_{x}B}^{Am})$ and $S(\rho_{\sigma_{z}B}^{Am})$ in Eq.~\eqref{A0} with $S(\rho_{\sigma_{x}B}^{Ph})$ and $S(\rho_{\sigma_{z}B}^{ph})$, then the
dynamics of the entropic uncertainty relation can be obtained exactly. However, the explicit expression is too complicated to present in the text. Nevertheless, numerical results indicate
that the dynamics of the entropic uncertainties (or their lower bounds) under phase damping channel with memory. In Fig. 2, we plot the uncertainties and their lower bounds of the observables $\sigma_{x}$ and $\sigma_{z}$ for initial state parameters $\vec{a}=\vec{b}=(0,0,0)$ and $C=\{1/2,-1/2,1/2\}$ as a function of $D$, respectively. One can find that the uncertainties or their low bonds of
two incompatible observables might be reduced under the influence of the phase damping noise with memory. The stronger the memory coefficient of channel $\mu$ is, the lower the uncertainties will be reduced. The reason
is that there is information flowing toward the system from noisy environment due to the memory effect, resulting in crease of our ability to precisely predict the outcomes of measurement so that the uncertainties reduce. In particularly, in the limit $\mu\rightarrow 1$, the uncertainties or their low bonds tends to a fixed value and is not effected by decoherence rate. Besides, we also display the purity of state $\rho_{AB}$ under phase damping channel with memory is dramatically anticorrelated with the entropic uncertainty relation in Fig. 2.

\subsection{Depolarizing channel with memory}
Depolarizing channel is also a unital channel, and the Kraus operators in the presence of memory are given in Table I, where $P_{0}=1-D$, $P_{1}=P_{2}=P_{3}=D/3$, with $D=1-\exp(-\gamma t)$. We resort to $\sigma_{x}$ and $\sigma_{z}$ as the incompatibility, the post measurement states are given as

\begin{widetext}
\begin{eqnarray}\label{Eq7}
\rho_{\sigma_{x}B}^{De}&=&\frac{1}{12}[3+(3-4D)b_{3}](|00\rangle\langle00|+|10\rangle\langle10|)+\frac{1}{12}[3-(3-4D)b_{3}](|01\rangle\langle01|+|11\rangle\langle11|)\nonumber\\
&+&\{\frac{1}{36}c_{1}[9-4D(6-4D)(1-\mu)](|00\rangle\langle11|+|01\rangle\langle10|)\nonumber\\
&+&\frac{1}{12}(3-4D)(b_{1}-i b_{2})(|00\rangle\langle01|+|10\rangle\langle11|)\nonumber\\
&+&\frac{1}{12}(3-4D)a_{1}(|01\rangle\langle11|+|00\rangle\langle10|)+H.C.\}\\
\rho_{\sigma_{z}B}^{De}&=&\frac{1}{36}\{9-(9-12D)(a_{3}+b_{3})+c_{3}[(3-4D)^2+8D(3-2D)\mu]\}|11\rangle\langle11|\nonumber\\
&+&\frac{1}{36}\{9-(9-12D)(a_{3}-b_{3})-c_{3}[(3-4D)^2+8D(3-2D)\mu]\}|10\rangle\langle10|\nonumber\\
&+&\frac{1}{36}\{9+(9-12D)(a_{3}-b_{3})-c_{3}[(3-4D)^2+8D(3-2D)\mu]\}|10\rangle\langle10|\nonumber\\
&+&\frac{1}{36}\{9+(9-12D)(a_{3}-b_{3})+c_{3}[(3-4D)^2+8D(3-2D)\mu]\}|10\rangle\langle10|\nonumber\\
&+&\{\frac{1}{12}(3-4D)(b_{1}-i b_{2})(|00\rangle\langle01|+|10\rangle\langle11|)+H.C.\}\\
\rho_{B}^{De}&=&\frac{1}{6}[3-(3-4D)b_{3}]|1\rangle\langle1|+\frac{1}{6}[3+(3-4D)b_{3}]|0\rangle\langle0|+
\{\frac{1}{6}(3-4D)(b_{1}-i b_{2})|0\rangle\langle1|+H.C.\}
\end{eqnarray}
\end{widetext}

\begin{figure}[htpb]
   \includegraphics[width=7.5cm]{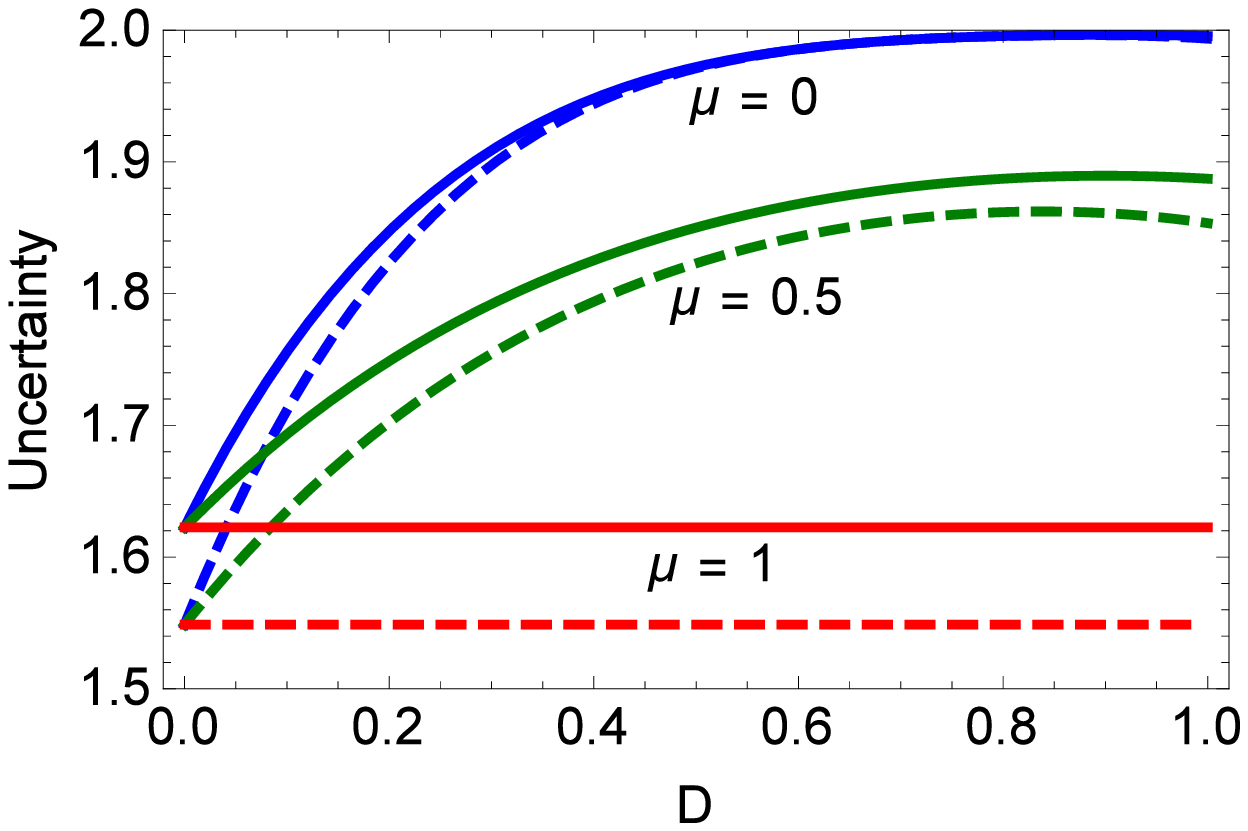}
    \includegraphics[width=7.5cm]{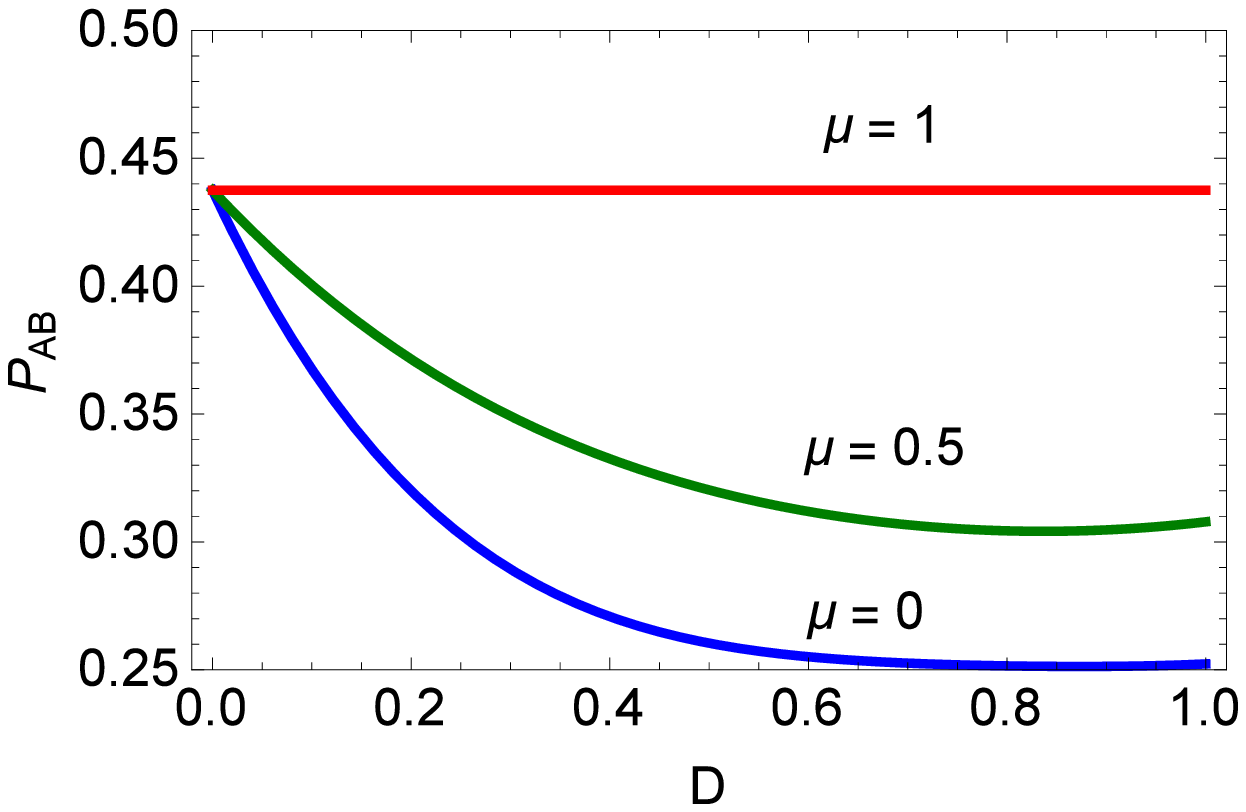}
 \caption{\label{Fig4}(Color online)The entropic uncertainties, their lower bounds and the purity of state $\rho_{AB}$ under depolarizing damping channel with memory for $D$ as decoherence rate. The solid lines represent the entropic uncertainties and the dashed lines represent the lower bounds of uncertainties.}
\end{figure}
Similarly, replacing the $S(\rho_{\sigma_{x}B}^{Am})$ and $S(\rho_{\sigma_{z}B}^{Am})$ in Eq.~\eqref{A0} with $S(\rho_{\sigma_{x}B}^{De})$ and $S(\rho_{\sigma_{z}B}^{De})$, the
dynamics of the entropic uncertainty relation can also be obtained. Here we numerically plot the uncertainties and their lower bounds of the observables $\sigma_{x}$ and $\sigma_{z}$ for initial state parameters $\vec{a}=\vec{b}=(0,0,0)$ and $C=\{1/2,-1/2,1/2\}$ as a function of $D$ in Fig.3, respectively. From this graph, one can find that the uncertainties suffer from depolarizing channel with memory have the same changing tendency with those under phase damping channel with memory. The stronger the memory coefficient of channel $\mu$ is, the lower the uncertainties will be reduced. This indicates the uncertainties or their low bonds of
two incompatible observables might be reduced under the influence of the depolarizing channel with memory. In particularly, in the limit $\mu\rightarrow 1$, the uncertainties or their low bonds is not effected by decoherence rate.

\section{Conclusion}  
In summary, we have investigated the entropic uncertainty relation under the influence of unital and nonunital noisy channels with memory. The effects of channels with memory such as amplitude damping channel(nonunitary), phase-damping and depolarizing channels(unitary), on dynamics of the entropic uncertainties (or their lower bounds) have been discussed in detail. Compared with the results obtained by Xu et al. Our results show that: Firstly, the entropic uncertainties (or their lower bounds) can be reduced regardless of in unital or nonunital channels due to channels with memory.  Secondly, the entropic uncertainties (or their lower bounds) subjecting to amplitude damping channel with memory will reduce at first and then lift with the memory coefficient of channel $\mu$ increasing. However, phase-damping and depolarizing channels with memory will always reduce the entropic uncertainties. The stronger the memory coefficient of channel $\mu$ is, the lower amount of the entropic uncertainties reduce.  Especially, in the limit of $\mu\rightarrow1$, the entropic uncertainties (or their lower bounds) are not influenced by decoherence of channle. Finally, the mechanism behind this phenomenon is also explored by using the purity of state. Thereby, we believe that our results might be helpful for insighting into
the dynamics behavior of the entropic uncertainty under open systems as well as its steering.

\acknowledgments
This research is supported by the Start-up Funds for Talent Introduction and Scientific Research of Changsha University 2015 (Grant No.SF1504) and the Scientific Research Project of Hunan Province Department of Education (Grant No.16C0134) and the Natural Science Foundation of Hunan Province (Grant No.
2017JJ3346), Key Laboratory of Low-Dimensional Quantum Structures and
Quantum Control of Ministry of Education (QSQC1403).

\label{app:eff-trans}

\end{document}